\documentclass[conference]{IEEEtran}
\IEEEoverridecommandlockouts
\usepackage{cite}
\usepackage{amsmath,amssymb,amsfonts}
\usepackage{placeins}

\usepackage{graphicx}
\usepackage{textcomp}
\usepackage{xcolor}
\def\BibTeX{{\rm B\kern-.05em{\sc i\kern-.025em b}\kern-.08em
    T\kern-.1667em\lower.7ex\hbox{E}\kern-.125emX}}

\usepackage{stfloats}
\usepackage{latexsym}
\usepackage{subcaption}
\usepackage[most]{tcolorbox}
\usepackage{color}
\usepackage{array}
\usepackage{adjustbox}
\usepackage[table]{xcolor}
\usepackage{url}
\usepackage{booktabs}
\usepackage{amsthm}

\usepackage{listings}
\usepackage{algorithm}
\usepackage{algpseudocode}

\usepackage{xspace}
\newcommand{\population}{P\xspace}
\newcommand{\evobudget}{B\xspace}
\newcommand{\pcrossover}{\ensuremath{p_c}\xspace}

\begin{document}


\title{EvoGPT: Leveraging LLM-Driven Seed Diversity to Improve Search-Based Test Suite Generation}

\author{
\IEEEauthorblockN{Lior Broide \qquad Roni Stern \qquad Argaman Mordoch}
\IEEEauthorblockA{
Faculty of Computer and Information Science\\
Ben-Gurion University of the Negev\\
broidel@post.bgu.ac.il \quad sternron@bgu.ac.il \quad mordocha@post.bgu.ac.il
}
}

\maketitle

\begin{figure*}
    \centering
    \includegraphics[width=0.75\textwidth]{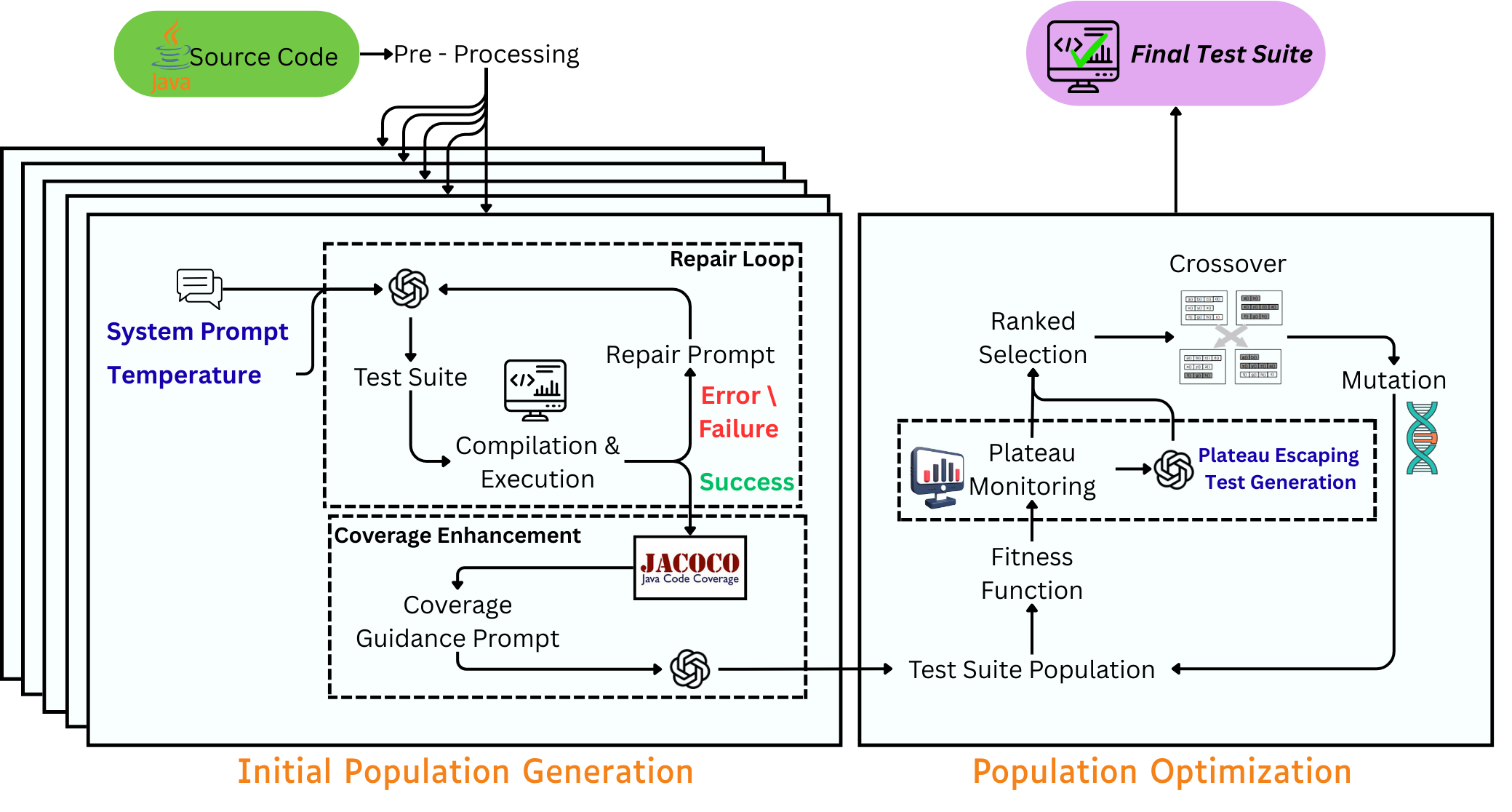}
    \caption{Overview of the EvoGPT system, illustrating the initial population generation process and the population optimization process. The former includes diverse system prompts and temperatures, a generation-repair loop, and a coverage enhancement step. The latter includes an evolutionary algorithm, search plateau monitoring, and an LLM-based plateau escaping test generation step that also uses diverse system prompts and temperatures.}
    \label{fig:diagram}
\end{figure*}

\begin{abstract}
Search-Based Software Testing (SBST) is a well-established approach for automated unit test generation, yet it often suffers from premature convergence and limited diversity in the generated test suites. 
Recently, Large Language Models (LLMs) have emerged as an alternative technique for unit test generation. 
We present EvoGPT, a hybrid test generation system that integrates LLM-based test generation with SBST-based test suite optimization. 
EvoGPT uses LLMs to generate an initial population of test suites, and uses an Evolutionary Algorithm (EA) to further optimize this test suite population. 
A distinguishing feature of EvoGPT is its explicit enforcement of diversity, achieved through the use of multiple temperatures and prompt instructions during test generation. 
In addition, each LLM-generated test is refined using a generation-repair loop and coverage-guided assertion generation. 
To address evolutionary plateaus, EvoGPT also detects stagnation during search and injects additional LLM-generated tests aimed at previously uncovered branches. Here too diversity is enforced using multiple temperatures and prompt instructions. 
We evaluate EvoGPT on Defects4J, a standard benchmark for test generation. The results show that EvoGPT  achieves, on average, a 10\% improvement in both code coverage and mutation score metrics compared to TestART, an LLM-only baseline; and EvoSuite, a standard SBST baseline. 
An ablation study indicates that explicitly enforcing diversity both at initialization and during the search is key to effectively leveraging LLMs for automated unit test generation.

\end{abstract}


\section{Introduction}

Software testing is a fundamental practice in software engineering, critical for ensuring the reliability, correctness, and maintainability of software systems~\cite{beck2003tdd}. Among its various forms, \textit{unit testing} serves as the first line of defense against bugs by verifying the behavior of individual program units in isolation. However, despite its importance, the process of manually designing high-quality unit tests remains labor-intensive and error-prone. 

\textit{Search-Based Software Testing} (SBST) is an extensively studied paradigm for automated test generation~\cite{harman2007searchbased,almulla2022learning, lukasczyk2022pynguin}. 
Techniques such as Evolutionary Algorithms (EAs), implemented in tools like \textit{EvoSuite}~\cite{fraser2012evosuite}, and \textit{DynaMosa}\cite{panichella2017automated}
systematically optimize test suites toward structural coverage objectives (e.g., branch and line coverage). 
While effective, SBST techniques are known to suffer from limitations such as coverage plateaus - where evolutionary search stagnates despite additional generations, and difficulty generating semantically diverse tests for complex behaviors.

In recent years, \textit{Large Language Models} (LLMs) have emerged as promising tools for automating unit test generation~\cite{chen2021codex}. Approaches such as \textit{ChatUniTest}~\cite{chen2024chatunitest}, \textit{TestART}~\cite{2024arXiv240803095G}, and \textit{A3Test}~\cite{xie2024a3test} demonstrate the ability of LLMs to produce syntactically correct, semantically coherent, and human-readable unit tests. 
These approaches leverage their natural language understanding and code generation capabilities to approximate human-level testing strategies. Nevertheless, LLM-based methods often suffer from limited exploration of complex input spaces, instability caused by hallucinations, and an inability to iteratively improve upon initial generations without external feedback~\cite{madaan2023self}. 

Motivated by the complementary strengths and weaknesses of SBST and LLM-based approaches, we propose \emph{EvoGPT}, a hybrid system that integrates diverse LLM-based test generation with evolutionary search-based test suite optimization. EvoGPT is illustrated in Figure~\ref{fig:diagram}. It generates a diverse initial population of test suites by querying LLMs under multiple system prompts and temperature settings, encouraging exploration across different execution paths and behavioral patterns. 
A generation-repair loop and a coverage enhancement step inspired by TestART~\cite{2024arXiv240803095G} further refines this initial population. 
Subsequently, EvoGPT uses an Evolutionary Algorithm (EA) that applies selection, crossover, and mutation operators to optimize a fitness function that combines structural coverage and mutation score. 

A key insight underlying EvoGPT is that the effectiveness of LLM-SBST hybridization critically depends on injecting \emph{semantically diverse} test suites at the \emph{right stages} of the evolutionary optimization process, rather than relying on a single-shot LLM initialization (i.e., generating a single LLM-based test population once, without further LLM involvement during search) or uniform intervention throughout the search. Accordingly, EvoGPT incorporates a \emph{plateau escape mechanism} inspired by CodaMOSA~\cite{lemieux2023codamosa}: when the evolutionary optimization process stagnates i.e., no significant fitness improvement is observed for a configurable number of generations, EvoGPT re-invokes multiple prompt and temperature LLM configurations to generate targeted test methods aimed at uncovered branches and mutations. These tests are injected into the current best test suite, enabling the search to escape local optima and continue improving.

EvoGPT is not the first test generation system to integrate search-based and LLM-based test generation methods~\cite{xiao2024optimizing,dakhama2025enhancing,lemieux2023codamosa,yang2025llm}. 
Most prior work focused on the question of \emph{when}  LLMs should be used in SBST, showing they are effective during initialization and after the search reached a plateau~\cite{xiao2024optimizing,lemieux2023codamosa}. 
In this work, we focus on a complementary question: \emph{how} to effectively generate and exploit semantically diverse LLM-generated candidates through prompt and temperature diversity, both at initialization and during plateau escaping, under equal evolutionary budgets.  

We evaluate EvoGPT on the Defects4J benchmark~\cite{just2014defects4j} and compare it against state-of-the-art SBST-only and LLM-only baselines, namely EvoSuite~\cite{fraser2012evosuite} and TestART~\cite{2024arXiv240803095G}. Our results show that EvoGPT significantly outperforms both baselines in line coverage, branch coverage, and mutation score of successfully generated tests. An ablation study further demonstrates the contribution of prompt diversity, evolutionary optimization, and plateau escape to EvoGPT’s overall effectiveness.


\noindent The main contributions of this paper are:
\begin{itemize}
    \item \textbf{A highly-effective SBST-LLM hybrid test generation system.} EvoGPT uses a novel method for combining search- and LLM-based test generation techniques to produce high-quality test suites.
    \item \textbf{Diversity inducing prompt and temperature configuration.} EvoGPT uses multiple structured prompt configurations and temperature settings to produce semantically diverse test cases. 
    Our ablation studies show empirically that this diversity greatly improves the quality of the resulting test suite. 

    \item \textbf{Empirical evaluation.} Experimental results using the Defects4J benchmark~\cite{just2014defects4j} show the benefits of using our EvoGPT system, demonstrating substantial improvements in code coverage and mutation testing over strong baselines, namely EvoSuite~\cite{fraser2012evosuite} and TestART~\cite{2024arXiv240803095G}, with gains driven by explicitly enforcing diversity in the generated test suites. 
\end{itemize}

This paper is organized as follows. 
Section~\ref{sec:background} provides necessary background and defines the test generation problem. 
Section~\ref{sec:evogpt} presents our EvoGPT test generation system. 
Section~\ref{sec:experimentalResults} details our experimental evaluation of EvoGPT and reports our empirical findings. 
Finally, Section~\ref{sec:conclusion} concludes and lists several directions for future research. 

\section{Background and Problem Definition}
\label{sec:background}

A \emph{unit test} is a self-contained function designed to validate the behavior
of a specific unit of source code, most commonly an individual method.
Executing a test typically involves invoking the target method under specific inputs and comparing the observed outcome against expected outputs through \emph{assertions}. The result of a test is often binary: a test either \emph{passes} or \emph{fails},\footnote{In practice, tests may also exhibit \emph{flakiness}, where outcomes vary across
executions due to non-determinism, environmental dependencies, or timing effects.
In this work, we focus on deterministic unit tests and consider only consistently
passing tests when evaluating test quality.} depending on whether all assertions hold. 
The objective of an automated unit test generation algorithm is to synthesize a suite of correct, high-quality, unit tests. 
Following prior work~\cite{2024arXiv240803095G}, we designate the \emph{public methods} of each public class under test as \textbf{focal methods}, which constitute the primary targets for test generation and evaluation.
This focus enables the testing effort to concentrate on actionable and externally observable behavior.
The following metrics are commonly used for evaluating the quality of a suite of unit tests~\cite{fraser2012evosuite,harman2007searchbased,jia2011survey}:
\begin{itemize}
    \item \textbf{Line Coverage of Correct Tests (LCCT)}: the percentage of source code lines in the focal methods that are executed at least once by a passing test. 
    \item \textbf{Branch Coverage of Correct Tests (BCCT)}: the percentage of conditional branches (e.g., if-else, switch cases) in the focal methods that are exercised in passed tests. This metric captures control-flow depth and is more sensitive to logical path diversity.
    \item \textbf{Mutation Score of Correct Tests (MSCT)}: the percentage of artificially injected faults (mutants) in the focal methods that are detected, i.e., cause at least one test to fail. This is a commonly used proxy for fault detection capability~\cite{fraser2012evosuite, jia2011survey}.
\end{itemize}
 

\subsection{Traditional Test Generation Methods.}
Many techniques have been proposed to automatically generate test suites. 
\emph{Random test generation} methods such as Randoop~\cite{pacheco2007randoop} sample inputs and assertions possibly applying heuristics for guidance. Assertions are inferred from observed runtime behavior, for example by recording returned values, object states, or thrown exceptions, and then checking for consistency in subsequent executions. These methods are computationally inexpensive, but often produce shallow tests with limited coverage and weak or missing assertions.

Search-Based Software Testing (SBST) methods, such as EvoSuite~\cite{fraser2012evosuite} apply evolutionary algorithms to optimize 
test quality metrics such as coverage and mutation score over multiple generations. 
These SBST methods begin by generating an initial population of random test suites, which are then iteratively refined using three core operators: \emph{selection}, which favors test suites with higher fitness for reproduction; \emph{crossover}, which combines parts of two parent test suites to create new ones; and \emph{mutation}, which introduces small random changes to individual tests such as modifying numeric or string literals within the test, removing assertions, etc. This evolutionary process gradually improves coverage and fault detection ability over time. 
Notably, applying evolutionary algorithms often requires implementing domain-specific fitness functions, cross-over and mutation operators. 
While EvoSuite and similar approaches can achieve high structural coverage, they struggle with assertion quality and test readability \cite{biagiola2025improving}.

Other methods for test generation require either additional knowledge of the software being tested or the ability to execute and instrument the tests to trace their execution. 
For example, metamorphic test generation assume as input a set of properties that relate the values of the tested units' inputs and outputs. 
These properties, known as metamorphic relations, guide the generation of follow-up tests
whose expected behavior can be inferred from previous executions~\cite{chen2018metamorphic}.
Test generation methods based on dynamic symbolic execution~\cite{chen2013state} execute a program
while simultaneously tracking symbolic expressions over input variables, capturing the
path conditions encountered during execution. 
These path constraints are then solved using constraint solvers (e.g., SMT solvers) to
generate new inputs that drive execution toward previously unexplored branches.

\subsection{LLM-based test generation methods.}
Recently, LLMs have been used to synthesize test cases based on natural language prompts or code structure. 
Early approaches relied on single shot prompts (e.g., ``Write a unit test for this method'') \cite{yuan2024chatgpt}, but more sophisticated techniques have since emerged. These include prompt engineering to guide assertion behavior, few-shot prompting to improve test structure \cite{chen2024chatunitest}, and iterative refinement using feedback such as stack traces or compilation errors\cite{2024arXiv240803095G}. 

For example, TestART~\cite{2024arXiv240803095G} combines LLM-based test generation with a feedback loop that repairs failing tests. If a generated test fails to compile or run, it captures the error and tries to fix it programmatically using a set of repair rules. 
If the error persists, it re-prompts the model with the error trace and original input, refining the test in multiple rounds. 
Meta's Automated Compliance Hardening (ACH)~\cite{harman2025mutation} is another example of an LLM-based test generation system. They focus on generating tests designed to find a specific type of bugs, using an LLM to generate mutations 
that simulate such bugs and then to generate tests that catch them. 


LLM-generated tests, however, often contain superficial validations and tend to overfit
syntactic patterns rather than exercising the underlying semantics of the code under
test~\cite{yuan2024chatgpt, chen2024chatunitest, testpilot2023}. Empirical comparisons have
shown that, while LLM-based approaches can achieve reasonable mutation scores, they
can lag behind search-based and symbolic execution techniques in structural coverage
and could be sensitive to the complexity of the code under test~\cite{abdullin2025test}. Moreover, 
the assertions generated by LLMs are frequently shallow, avoid exercising deep program
logic, and fail to substantially improve test quality in the absence of external feedback.



\subsection{LLMs and Evolutionary Algorithms}
Recent works have begun to explore how LLMs and
Evolutionary Algorithms (EAs) can complement one another for test generation and for other tasks. 
Broadly, LLMs can (i) act as evolutionary operators by proposing
candidate solutions like seeding the initial population, (ii) provide semantic guidance that helps EAs escape
local optima, or (iii) encode problem structure that would otherwise be
hard to capture with hand-designed heuristics. 
Liu et al.~\cite{liu2024large} showed that an LLM can be prompted to
perform selection, crossover, and mutation directly, demonstrating that
LLMs can function as zero-shot evolutionary optimizers. 
Wang et al.~\cite{wang2025multi} further extended this idea to
multi-agent settings where specialized LLM agents collectively drive an
evolutionary search process. 

Wu et al.~\cite{wu2024evolutionary} provide a taxonomy of these
approaches, distinguishing between \emph{LLM-enhanced EAs} 
(using LLMs to support or improve evolutionary search), 
\emph{EA-enhanced LLMs} (using evolutionary strategies to tune or guide
LLMs), and tightly integrated hybrid systems. 
EvoGPT fits into the category of \emph{LLM-enhanced EAs}. 
Rather than replacing the evolutionary loop, we use LLMs to generate a
semantically diverse initial population and to provide guided injections
when the search plateaus. 
This combination retains the strengths of traditional SBST
while exploiting the generative and semantic capabilities of modern LLMs.

\subsection{LLMs and EAs for Automated Test Generation}
The integration of LLMs and EAs specifically for automated test generation has recently gained momentum, motivated by the complementary strengths of semantic
reasoning from LLMs and systematic exploration from EAs.
Xiao et al.~\cite{xiao2024optimizing} investigate three places where LLM assistance can be used in an EA for test generation: 
(i) generating the initial population of test suites,
(ii) injecting additional test cases during the evolutionary process to target uncovered code,
and (iii) introducing new candidates when the search stagnates.
Their experimental results show that LLM-based assistance yields the strongest benefits
during the initialization phase and when applied after moderate stagnation, whereas
overly frequent intervention provides diminishing returns.

Another influential system in this space is \emph{CodaMosa} by Microsoft
Research~\cite{lemieux2023codamosa}. 
CodaMosa is an EA-based software test generation system that uses an LLM to escape structural coverage plateaus.
Rather than relying solely on mutation and crossover, CodaMosa periodically queries an LLM 
to synthesize new tests aimed towards areas of the tested program that are less covered. 
We implemented a similar plateau-escaping mechanism in our EvoGPT. 


Yang et al.~\cite{yang2025llm} created pytLMtester, a test generation system that extends DynaMosa~\cite{panichella2017automated}
for dynamically typed languages using LLMs. 
They run an EA to generate tests and use LLMs to infer object types, repair failed tests, and generate mutations. 
Hybrid methods for automated test generation have also been explored in other domains.
SearchSYS~\cite{dakhama2025enhancing} uses LLMs to generate high-level test templates that
guide a search-based fuzzer in system-level simulators, showing that LLM-derived structure
can improve bug-finding effectiveness.


EvoGPT builds upon and extends several of the aforementioned hybrid efforts. Unlike most prior work, it does not rely on a single LLM invocation to generate tests, but employs a diverse set of LLM configurations that differ in their prompts and temperatures. This generates initial test suites that have distinct and diverse structural and semantic characteristics, which we show to be crucial in designing an effective LLM-SBST hybrid test generation system.





\section{The EvoGPT System}
\label{sec:evogpt}
In this section, we describe the EvoGPT system, which combines LLMs and EA to generate and optimize effective and diverse test suites. 
Figure~\ref{fig:diagram} provides a high-level illustration of EvoGPT. 
EvoGPT consists of two main processes: (1) an \emph{Initial Population Generation} process, which generates an initial set of test suites, and (2) a \emph{Population Optimization} process, which employs an EA to iteratively refine the generated test suites and outputs a single optimized test suite.



The initial population generation process instantiates five distinct LLM agent configurations, each defined by a different prompting strategy (e.g., focusing on edge cases, deep object chains, or creative assertion styles) as well as different temperature values. 
Each agent is queried asynchronously five times, resulting in a total of 25 diverse and structurally unique test suites. 
During this generation process, a lightweight \emph{repair loop} inspired by TestART~\cite{2024arXiv240803095G}, is applied to each test suite: when a generated test fails due to compilation or runtime errors, the system captures the stack trace and attempts to fix it. This helps improve test validity and robustness early in the pipeline. 
Successful test suites go through an additional \emph{coverage enhancement} step, which involves identifying uncovered branches in the code and calling an LLM to attempt to add tests that cover them. 
The resulting 25 test suites are passed to EvoGPT's population optimization process, in which an EA iteratively evolves this population by applying selection, crossover, and mutation operations guided by a fitness function.  
Inspired by CodaMosa~\cite{lemieux2023codamosa}, when EvoGPT detects stagnation in the EA, it injects additional tests 
to help escape local optima. 
Here as well, multiple LLM agents are used to generate these injected tests, to encourage diversity. 
Next, we describe the details of the EvoGPT main processes: initial population generation process and population optimization. 


\subsection{Initial Population Generation}

Prior to starting the initial population generation process, EvoGPT performs a lightweight pre-processing on the source code to reduce token complexity and context window. 
Specifically, we remove excessive or lengthy inline comments, documentation blocks, and unreachable code segments that could overload the model's context window and negatively affect generation quality~\cite{gao-etal-2024-insights}. 
Then, we generate 25 test suites for a given class by instantiating five LLM agents and having each agent generate five test suites. 

Importantly, each LLM agent is configured with a different temperature setting and a corresponding system prompt. 
The temperature value is used to control the stochasticity of the language model output. 
Prior research has shown that varying temperature values leads to a broader range of generated outputs, exposing different behavioral patterns and increasing semantic variance across samples~\cite{agarwal2024temperature}. 
To complement this, each agent is given a custom system prompt, which directs it to approach the test suite generation task from a slightly different perspective (e.g., focusing on execution paths, edge values, or assertion robustness). 
Table~\ref{tab:agent-config} lists for each of the five types of LLM agents we used, their temperature value, a qualitative description of their prompt (column ``Prompt objective''), and intended purpose (column ``Purpose''). 
The exact system prompts used for each LLM agent are included in the provided code. 
This diversity promotes both semantic variability and robustness in the generated tests. 
We empirically verified that this multi-prompt configuration produces semantically diverse
initial test suites. A detailed analysis of the diversity metrics and statistical tests is presented in Section~\ref{sec:experimentalResults}.

\begin{table*}[h!]
\centering
\caption{LLM agent configuration: temperature and prompt strategy paired to encourage diversity and robustness.}
\begin{tabular}{@{} p{0.06\linewidth} p{0.14\linewidth} p{0.74\linewidth}}
\toprule
\textbf{Agent} (Temp.) & \textbf{Purpose}  & \textbf{Prompt objective}\\
\midrule
A1 \newline(temp.=0.3) & Standard unit testing & 
\begin{scriptsize}
1. Cover as many branches as possible 
in the "focal method" (Branch Coverage).  \newline
2. Write meaningful assertions. \newline
3. If you suspect a section of the code to be vulnerable to potential mutations, write a meaningful assertion or whole test to catch it.
\end{scriptsize}
\\
\midrule
A2 \newline(temp.=0.6) & Tests with diverse assertions. 
& 
\begin{scriptsize}
1. Cover all reachable branches, but prioritize 
writing multiple strong and diverse assertions 
that verify: \newline
   - Return values \newline
   - Internal object state (via getters) \newline
   - Side effects on collections or parameters \newline
2. Maximize the use of assertions in every test case, checking both normal and edge values. \newline
3. If you suspect a section of the code to be vulnerable to mutations, write a meaningful assertion or whole test to catch it. 
\end{scriptsize}
\\
\midrule
A3 \newline(temp.=0.8) & “Try hard” creative agent, aiming for path and value exploration & 
\begin{scriptsize}
1. Focus on inputs and execution paths that are semantically meaningful, 
   surprising, or likely to violate developer assumptions. \newline
2. Actively explore edge values, boundary conditions, and atypical input 
   combinations (e.g., empty, extreme, inconsistent, or contradictory values). \newline
3. Analyze the method for sections that may be prone to logic errors, misuse of conditionals, or potential edge case failures. Write full test cases specifically to **expose potential bugs**, even if coverage is low. 
\end{scriptsize}
\\
\midrule
A4 \newline(temp.= 0.5) & Focus on edge conditions, detecting boundary cases & 
\begin{scriptsize}
1. Focus on edge cases and boundary values that may trigger exceptional paths. For example: empty strings, null values, min/max ints, empty arrays, single-element lists, etc. \newline
2. Each test should target one edge condition at a time. 
\end{scriptsize}
\\
\midrule
A5 \newline(temp.=0.4) & Uses long object chains, aiming for deeper semantics & 
\begin{scriptsize}
1. Cover **as many branches as possible** in the focal method using the fewest number of test cases.\newline
2. Write exactly one assertion per test method — pick the one that validates the key condition.\newline
3. Avoid redundant or overly detailed checks unless necessary for coverage.
\end{scriptsize}
\\
\bottomrule
\end{tabular}%
\label{tab:agent-config}
\end{table*}



\paragraph{Generation-Repair Loop}
After generating the initial set of tests, each LLM agent performs a \emph{generation-repair loop} where it iteratively refines its test case in response to runtime feedback such as compilation errors or failing assertions. 
Such a generation-repair process was shown to enhance test validity and robustness in prior LLM-based systems such as ChatUniTest~\cite{chen2024chatunitest}, TestART~\cite{2024arXiv240803095G} and TestPilot~\cite{testpilot2023}. In our system, this loop leverages a simplified version of TestART’s repair strategy: when a generated test fails with a runtime error or stack trace, EvoGPT programmatically extracts the stack trace and invokes a \emph{minimal repair heuristic} to remove or correct the failing statement. 
Our minimal repair heuristic directly fixes common runtime and compilation errors. 
For instance, when an import error is raised, the repair loop programmatically traverses through the source code's repository in order to fetch the right import paths.
If our minimal repair heuristic fails to fix the generated test, we call the LLM with a repair prompt along with the stack trace that define the runtime or compilation error. The exact repair prompt is also included in our code. 
Using a programmatic, targeted, template-based repairs before calling an LLM-based repair has proved to be more stable and cost-effective for improving test pass rates than only using an LLM-based repair~\cite{2024arXiv240803095G}


This repair loop is repeated at most four times, bounding the accumulation of error traces
and code context passed to the LLM, thereby limiting prompt growth and reducing the risk
of context window saturation~\cite{gao-etal-2024-insights}.
If errors in the tests persist, the test methods responsible for these errors are automatically removed from the test suite.

\paragraph{Enhancing Coverage}
Once a valid test suite is synthesized, it is compiled and executed using JaCoCo,\footnote{\url{www.jacoco.org/jacoco}} a Java Code Coverage tool, which provides detailed line and branch coverage information. The resulting coverage report is parsed and passed to the next module: the Coverage Analysis Agent. This agent uses the same temperature as the original generation agent to maintain behavioral consistency. 
It is then instructed to generate additional complementary tests, aimed specifically at covering the missed branches and lines reported in the coverage report. This strategy was found valuable for enhancing code coverage, as demonstrated in TestART~\cite {2024arXiv240803095G}.
Finally, the tests produced by both the initial generation agent and its corresponding coverage enhancement process are merged into a single composite test suite. This forms one member of the initial population in our evolutionary algorithm.

\subsection{Population Optimization}

\begin{algorithm}[ht]
\caption{EvoGPT Population Optimization Process}
\label{alg:evolution}
\begin{algorithmic}[1]
\Require $\population, \evobudget, \pcrossover, \alpha, \tau, k$
\Ensure Highest-fitness test suite $T^*$
\State $s \gets 0; j\gets 0$; \Comment{Stagnation and injection counters}
\State $f_{\text{best}} \gets \max_{T \in P} \textsc{Fitness}(T)$
\State $T_{\text{init}} \gets \arg\max_{T \in P} \textsc{Fitness}(T)$ 
\For{$g \gets 1$ \textbf{to} \evobudget}
    \State $T^*  \gets \arg\max_{T \in P} \textsc{Fitness}(T)$
    \If{$f_{\text{best}} = 100$} \Comment{Perfect coverage and mutation}
        \State \Return $T^*$ \label{alg:line-early-terminate}
    \EndIf
    \State $(T_1, T_2) \gets \textsc{RankedSelection}(P)$ \label{alg:line-selection}
    \State $(O_1, O_2) \gets \textsc{Crossover}(T_1, T_2,\pcrossover)$ \label{alg:line-crossover}
    \State $O_1 \gets \textsc{Mutate}(O_1)$ \label{alg:line-mutate1}
    \State $O_2 \gets \textsc{Mutate}(O_2)$ \label{alg:line-mutate2}
    \If{$(O_1, O_2)$ is better than $(T_1, T_2)$ \label{alg:line-is-better}}
        \State $P \gets P \setminus \{T_1,T_2\} \cup \{O_1, O_2\}$  \label{alg:line-replace}
    \EndIf
    \State $f_{\text{curr}} \gets \max_{T \in P} \textsc{Fitness}(T)$\label{alg:line-check-stagnation}
    \If{$f_{\text{curr}} - f_{\text{best}} \geq \alpha$}
        \State $s \gets 0$; \quad $f_{\text{best}} \gets f_{\text{curr}}$
    \Else
        \State $s \gets s + 1$
    \EndIf
    \If{$s \geq \tau$ \textbf{and} $j < k$}
        \State $T^* \gets \arg\max_{T \in P} \textsc{Fitness}(T)$
        \State \Call{PlateauEscape}{$T^*$, $P$} \label{alg:line-llm-injection} \Comment{Inject new tests}
        \State $s \gets 0$; \quad $j \gets j + 1$
        \State $f_{\text{best}} \gets \max_{T \in P} \textsc{Fitness}(T)$\label{alg:line-end-injection-phase}
    \EndIf
\EndFor
\If{$T_{\text{init}} \notin P$}
    \State $P \gets P \cup \{T_{\text{init}}\}$ \Comment{Ensure best initial is preserved}
\EndIf
\State \Return $\arg\max_{T \in P} \textsc{Fitness}(T)$ \label{alg:line-return}
\end{algorithmic}
\end{algorithm}

To optimize the test suite population generated by EvoGPT's initial population generation process, we employ a customized EA. 
This EA takes as input a population of test suites $P$ and an evolutionary budget \evobudget, specifying the maximal number of EA iterations allowed. 
It uses a dedicated fitness function to evaluate test suites and applies the common EA operators of \emph{selection}, \emph{crossover}, and \emph{mutation}. Our EA also includes a plateau escaping operator that is applied on-demand. 
Eventually, it returns the test suite $T^*$ that maximizes our fitness score function. 
Next, we describe our implementation of these EA elements, and provide a complete pseudo-code specifying how they work together.

\subsubsection{Fitness Score}

The fitness function plays a central role in guiding the EA toward higher-quality solutions through what is known as \emph{evolutionary pressure} --- the selective process by which better-performing candidates are favored for reproduction, gradually improving the population over successive generations. 
In EvoGPT, the primary goal is to generate tests that are effective at exposing potential faults. 
To this end, EvoGPT uses the following fitness function, which is a linear combination of MSCT, BCCT and LCCT:
\begin{equation}
\begin{adjustbox}{max width=\columnwidth}
$\texttt{fitness} = 0.3 \cdot \texttt{BCCT} + 0.2 \cdot \texttt{LCCT} + 0.5 \cdot \texttt{MSCT}$
\end{adjustbox}
\end{equation}



These specific weights were chosen based on our guiding priorities: MSCT receives the highest weight (0.5) to emphasize fault detection; BCCT (0.3) is prioritized over LCCT (0.2) as it better captures logical path diversity. 
In preliminary experiments, we evaluated several alternative weight configurations
and found that this formulation provided the most stable convergence behavior and 
best structural and mutation-based metrics.


\subsubsection{Selection}

EvoGPT employs ranked selection 
to choose parent test suites for reproduction. The
population is sorted by our fitness function, and selection probabilities are assigned according to
rank rather than raw fitness values. 
Ranked selection is a common choice in EAs~\cite{whitley1989genitor,goldberg1991comparative,baker2014adaptive}, as it helps
control selection pressure and reduce the risk of premature convergence without requiring
problem-specific tuning.

\subsubsection{Crossover}
Given two parent test suites, offspring are created with the crossover operation by recombining test methods, helper methods,
and inner classes. 
In our implementation, we applied crossover at the test-suite level with probability $0.8$.
One offspring inherits approximately 80\% of its structure from the
first parent and 20\% from the second, with the inverse applied to the other offspring.

\subsubsection{Mutation}
An additional operator used in EAs is \textbf{Mutation}. Mutation introduces random variations into individuals, ensuring continued diversity and helping escape local optima.
We implement mutation operators 
based on those used by EvoSuite \cite{fraser2012evosuite}, applying structural transformations to test methods.  
Our mutation strategy employs three operators, each selected uniformly at random: 
\begin{enumerate}
    \item \textbf{Statement Deletion}: Removes a random non-assertion statement from the test method, simplifying the test while preserving its assertions and validation semantics.
    \item \textbf{Primitive Modification}: Modifies numeric or string literals within the test. Numeric values undergo small perturbations (i.e., increments or decrements), while strings are mutated through character addition, removal, or case changes. This explores neighboring regions of the input space.
    \item \textbf{Assertion Removal}: Removes a randomly selected assertion from the test methods containing multiple assertions when always preserving at least one test method, following EvoSuite's assertion minimization approach to identify assertion sets.
\end{enumerate}
In our implementation, each test method has a uniform mutation probability, where a mutation is added to given test suite with probability $\frac{1}{N_{\texttt{tests}}}$,
where $N_{\texttt{tests}}$ is the number of test methods in the mutated test suite. 
Thus, on expectation a single mutation is performed in every test suite. This conservative rate is inspired by common evolutionary computation practice, and aims to preserve test structure.

\subsubsection{Plateau Escape via LLM Injection}
\label{sec:plateau-escape}
EAs can become trapped in local optima or plateau, where standard genetic operators fail to significantly improve fitness despite continued iterations. Inspired by CodaMosa \cite{lemieux2023codamosa}, EvoGPT incorporates an \emph{LLM Injection} mechanism to escape such plateaus. 
The five LLM agents mentioned in previous section are queried in parallel. Each agent receives as input the current test suite context (imports, fields, and setup methods), the source code under test, and detailed coverage feedback including the specific uncovered branches and mutants identified. 
Rather than generating complete test suites, each LLM agent only adds new test methods within an existing test suite (i.e., only \texttt{@Test} methods). 
These new tests are validated and injected into the current best individual in the population. 
This design allows EvoGPT to enrich promising test suites with targeted behaviors without disrupting their existing structure, enabling the EA to escape local optima when stagnating, and resume effective exploration.

\subsubsection{Pseudo-Code}

Algorithm~\ref{alg:evolution} lists a pseudo-code of the entire EvoGPT population optimization process. 
It accepts as input as 
an initial population of test suites \population; the evolutionary budget \evobudget;
the crossover probabilities \pcrossover; 
the fitness improvement threshold $\alpha$; 
the stagnation threshold $\tau$; 
and the maximal number of attempts to inject tests to escape a plateau per generation $k$. 
In our implementation, $|\population|=25$, $\evobudget=25$, $\pcrossover=0.8$,
$\alpha=0.5$, 
$\tau=5$, 
and $k=3$. 
Throughout its execution, EvoGPT maintains the best fitness score found so far $f_{best}$. 
Every iteration starts by checking if the best test suite found so far has a perfect fitness score, which means that the code coverage and mutation scores of correct tests reached 100\%.
If this occurs, EvoGPT terminates early and returns this test suite (line~\ref{alg:line-early-terminate}). 
Otherwise, EvoGPT uses a ranked selection process to select the best two test suites $T_1$ and $T_2$. Then it generates a new pair of test suites $O_1$ and $O_2$ by performing a combination of crossover (line~\ref{alg:line-crossover}) and mutation (lines~\ref{alg:line-mutate1}-\ref{alg:line-mutate2}) operators. Each of these operators is applied stochastically: the crossover operator with probability \pcrossover and each mutation with probability $\frac{1}{N_{tests}}$. 
If the resulting pair of test suites include a test suite better than the ones used to generate it then $O_1$ and $O_2$ replace $T_1$ and $T_2$ in the population $P$ (line~\ref{alg:line-replace}). 
We checked if $O_1$ and $O_2$ are better than $T_1$ and $T_2$ (line~\ref{alg:line-is-better}) by computing the max of their fitness scores, breaking ties in favor of smaller test suites. 
That is, the pair $(O_1,O_2)$ was chosen over $(T_1, T_2)$ if 
$\max(\textsc{Fitness}(O_1), \textsc{Fitness}(O_2))>\max(\textsc{Fitness}(T_1), \textsc{Fitness}(T_2))$ if their fitness is the same and 
$|O_1| + |O_2| \leq |T_1| + |T_2|$.

Subsequently, EvoGPT monitors the evolutionary progress to check if a plateau has been reached, i.e., how many generations since a significant improvement in the fitness score has been observed.  
If more than $\tau$ generations has passed without such an improvement, then the plateau escaping mechanism mentioned in Section~\ref{sec:plateau-escape} is invoked (line~\ref{alg:line-llm-injection}). 
After the evolutionary budget has been spent (or a perfect test suite has been found), EvoGPT returns the best test suite found in the current population (line~\ref{alg:line-return}).


\section{Experimental Results}
\label{sec:experimentalResults}
We conducted an experimental evaluation of EvoGPT using Defects4J~\cite{just2014defects4j}, a benchmark of real-world Java projects with reproducible bugs and accompanying test suites. 
Defects4J is commonly used in the automated testing literature due to its diversity, complexity, and grounding in real software engineering practice. 
Table~\ref{tab:dataset} presents the Defects4J projects used in our experiments. 
Columns ``Identifier'', ``Project name'', ``Version'', and ``Focals'' represent 
the identifier of the repository under test, 
its name, 
the version of the code being tested, 
and the number of focal methods extracted from each project, respectively. 
Overall we experimented with all 17 repositories in the Defects4J benchmark and extracted at least 137 focal methods for each repository. 

\begin{table}[ht]
\centering
\caption{Dataset overview: Projects with the evaluated version number and the number of focal methods.}
\label{tab:dataset}
\resizebox{0.9\columnwidth}{!}{%
\begin{tabular}{lllc}
\toprule
\textbf{Identifier} & \textbf{Project name}  & \textbf{Version} & \textbf{Focals} \\ \midrule
Cli & Commons-Cli            & 1.6.0            & 177                    \\
Codec & Commons-Codec          & 1.17.1           & 563                    \\
Collections & Commons-Collections    & 4.5.0            & 2997                   \\
Compress & Commons-Compress       & 1.27.1           & 2288                   \\
Csv & Commons-Csv            & 1.10.0           & 137                    \\
JxPath & Commons-JXPath         & 1.3              & 998                    \\
Lang & Commons-Lang           & 3.17.0           & 1728                   \\
Math & Commons-Math           & 3.6.1            & 5136                   \\
Gson & gson                   & 2.11.0           & 378                    \\
JacksonCore & Jackson-Core           & 2.18.2           & 1782                   \\
JacksonDatabind & Jackson-Databind       & 2.18.2           & 5295                   \\
JacksonXml & Jackson-Dataformat-XML & 2.18.2           & 382                    \\
Chart & JFreeChart             & 1.5.5            & 6060                   \\
Jsoup & jsoup                  & 1.18.3           & 836                    \\
Mockito & mockito                & 5.14.2           & 1262                   \\
Time & Joda-time                   & 2.12.7           & 2824                   \\
Closure & Closure-compiler                & v20231112        & 4723                   \\ 
\bottomrule
\end{tabular}%
}
\end{table}

\begin{table}
\caption{Comparison across all evaluated projects. 
The columns EG, ES, and T show results for EvoGPT, EvoSuite, and TestART, respectively.}
\resizebox{\columnwidth}{!}{
\begin{tabular}{@{}l|lll|lll|lll@{}}
\toprule
                & \multicolumn{3}{c}{\textbf{LCCT}}            & \multicolumn{3}{c}{\textbf{BCCT}}         & \multicolumn{3}{c}{\textbf{MSCT}}         \\ 
  \textbf{Project}                      & EG      & ES    & T & EG      & ES    & T & EG      & ES    & T \\ \midrule
Closure       & \textbf{90} & \textbf{90} & 85      & \textbf{89} & 86       & 81      & \textbf{84} & 71       & 77      \\
Cli            & \textbf{96} & 90          & 87      & \textbf{93} & 89       & 87      & \textbf{92} & 80       & 86      \\
Codec          & \textbf{96} & 80          & 83      & \textbf{93} & 77       & 81      & \textbf{85} & 64       & 78      \\
Collections    & \textbf{95} & 79          & 81      & \textbf{91} & 74       & 83      & \textbf{92} & 68       & 79      \\
Compress       & \textbf{84} & 72          & 76      & \textbf{83} & 69       & 72      & \textbf{80} & 62       & 71      \\
Csv            & \textbf{96} & 78          & 81      & \textbf{94} & 65       & 79      & \textbf{97} & 58       & 77      \\
JxPath         & \textbf{90} & 77          & 76      & \textbf{87} & 70       & 73      & \textbf{82} & 63       & 71      \\
Lang           & \textbf{94} & 90          & 88      & \textbf{94} & 87       & 84      & \textbf{85} & 75       & 83      \\
Math           & \textbf{96} & 86          & 86      & \textbf{93} & 80       & 81      & \textbf{91} & 69       & 80      \\
Gson                   & \textbf{93} & 80          & 82      & \textbf{91} & 80       & 79      & \textbf{85} & 62       & 76      \\
JacksonCore           & \textbf{89} & 77          & 79      & \textbf{90} & 76       & 75      & \textbf{83} & 69       & 79      \\
JacksonDatabind       & \textbf{86} & 78          & 78      & \textbf{81} & 75       & 77      & \textbf{84} & 65       & 76      \\
JacksonXml & \textbf{90} & 83          & 87      & \textbf{91} & 79       & 82      & \textbf{89} & 70       & 80      \\
Chart             & \textbf{91} & 87          & 87      & \textbf{91} & 85       & 84      & \textbf{87} & 76       & 81      \\
Time              & \textbf{91} & 85          & 83      & \textbf{89} & 84       & 81      & \textbf{84} & 72       & 78      \\
Jsoup                  & \textbf{92} & 86          & 85      & \textbf{92} & 82       & 82      & \textbf{92} & 71       & 80      \\
Mockito                & \textbf{92} & 88          & 87      & \textbf{88} & 84       & 83      & \textbf{83} & 73       & 78      \\ \midrule
TOTAL                  & \textbf{92} & 83          & 83      & \textbf{90} & 79       & 80      & \textbf{87} & 69       & 78      \\ \bottomrule
\end{tabular}
}
\label{tab:framework-comparison}
\end{table}

We compared EvoGPT to two baselines: EvoSuite~\cite{fraser2012evosuite}, a widely used EA-based test generation system, and TestART~\cite{2024arXiv240803095G}, a recent state-of-the-art system for LLM-based unit test generation.
We chose TestART since it has been shown to outperform previous LLM-based approaches such as ChatUniTest~\cite{chen2024chatunitest}.
We did not include hybrid LLM-SBST approaches as baselines in our experiments as CodaMosa~\cite{lemieux2023codamosa} and pytLMtester~\cite{yang2025llm} are specifically designed for Python, a dynamically typed language, and SearchSYS~\cite{dakhama2025enhancing} targets system-level simulators rather than unit-level Java programs.

To measure the quality of the generated tests, we used the three test quality metrics mentioned above -- LCCT, BCCT, and MSCT --  restricted to passing tests\footnote{We omit tests that cannot be repaired in the repair loop (I.e fail in either runtime or compilation), similar to EvoSuite and TestART.} and focal method coverage. 
We measured MSCT based on the mutant analysis of PITest (version 1.19.0), while coverage is computed with JaCoCo (version 0.8.12). 
The LLM used in our experiments for EvoGPT and TestART is gpt-4o-mini. 
For Evosuite, we set the population size to 25 and the maximum number of generations to 25, to ensure a fair comparison with EvoGPT, which also starts with an initial population of 25 and runs for at most 25 generations. 
All other EvoSuite's parameters (including assertion minimization, combined coverage criteria, selection operators, etc) were left unchanged as the default command-line configuration provided by the tool.  

\subsection{Comparison Against Baselines}

Table~\ref{tab:framework-comparison} presents the average LCCT, BCCT, and MSCT scores across our benchmark projects. Per project values are the average values for all focal classes tested in the project.
The columns ``EG'', ``ES'', and ``T'' represents the results for EvoGPT, EvoSuite, and TestART, respectively.
The columns ``LCCT'', ``BCCT'' and ``MSCT'' represent the line, branch and mutation score coverages of the correct tests. 
The results clearly show that
EvoGPT consistently outperforms both baselines across all metrics. It achieves the highest total average MSCT (87), improving over TestART (78) and EvoSuite (69). 
Similarly, EvoGPT leads in both LCCT (92) and BCCT (90). 
These results suggest that EvoGPT is able to effectively integrate the benefits of LLM-based and search-based test generation techniques, and that its diverse initial population generation, and the plateau-escape strategies are effective. 



\begin{table}[htbp]
\centering
\caption{Wilcoxon signed-rank test results comparing EvoGPT with TestART and EvoSuite across evaluation metrics.}
\begin{tabular}{llcc}
\toprule
\textbf{Metric} & \textbf{Comparison} & \textbf{p-value} & \textbf{$\delta$} \\
\midrule
LCCT & EvoGPT vs TestART   & $<$0.001 & 0.76 \\
 & EvoGPT vs EvoSuite  & $<$0.001 & 0.79 \\
BCCT & EvoGPT vs TestART   & $<$0.001 & 0.79 \\
 & EvoGPT vs EvoSuite  & $<$0.001 & 0.82 \\
MSCT & EvoGPT vs TestART   & $<$0.001 & 0.75 \\
 & EvoGPT vs EvoSuite  & $<$0.001 & 0.98 \\
\bottomrule
\end{tabular}
\label{tab:significance}
\end{table}

Next, we conducted Wilcoxon signed-rank tests \cite{wilcoxon1992individual} 
to validate that the results above are statistically significant. 
We compared EvoGPT against TestART and EvoSuite across all of our dataset projects ($n=17$ projects) and evaluation metrics. We chose the Wilcoxon test as it is a non-parametric alternative appropriate for paired samples without assuming normal distribution, following recommendations for empirical software engineering studies~\cite{arcuri2011practical}.

The tests were performed over per-project results for LCCT, BCCT and MSCT across all 17 benchmark projects. 
We report also Cliff's delta ($\delta$), which is the ratio of cases where one algorithm was better than the other minus the ratio of the cases where the other algorithm was better~\cite{cliff1993dominance}. 
This is a known way to measure effect size, where $|\delta| \geq 0.474$ indicates a large effect.

The results are displayed in Table~\ref{tab:significance}.
The columns ``Metric'', ``Comparison'', ``p-value'' and ``$\delta$'' represent the evaluated metric (LCCT, BCCT or MSCT); the compared systems; the p-value score achieved; and Cliff's delta, respectively.
EvoGPT \textit{significantly} outperforms both baselines across all metrics ($p < 0.001$). 
All comparisons exhibit large effect sizes ($\delta \geq 0.75$), with the MSCT improvement over EvoSuite showing a very large effect ($\delta = 0.98$). 
These findings confirm that EvoGPT's performance gains - averaging +7.4\% over TestART and +11.6\% over EvoSuite - are statistically robust and practically significant. 
The consistent advantage across all 17 projects and three metrics validates the reliability of our approach.

\subsection{Sensitivity Analysis and Ablation Studies}
To better understand the contribution of individual components in EvoGPT, we conduct a series of parameter sensitivity analysis and ablation studies. These experiments systematically disable or modify core aspects of EvoGPT to assess their impact on test effectiveness. 

\subsubsection{Impact of Initial Population Size}
We varied the number of initial test population size to be 10, 15, 20, 25, and 30, while keeping all other parameters fixed. 
This was done by having each of our five temperature and prompt configurations run multiple times (2, 3, 4, 5, and 6 to generate 10, 15, 20, 25, and 30, respectively). 
Table~\ref{tab:initial_population_sensitivity} shows the LCCT, BCCT, and MSCT results (in each row) with different initial population sizes (in each column).

\begin{table}[ht]
\centering
\caption{EvoGPT performance per initial population size.}
\label{tab:initial_population_sensitivity}
\begin{tabular}{l c c c c c}
\toprule
\textbf{Initial Population} & 10 & 15 & 20 & 25 & 30 \\
\midrule
\textbf{MSCT} & 0.75 & 0.78 & 0.82 & 0.87 & 0.87 \\
\textbf{LCCT} & 0.78 & 0.85 & 0.88 & 0.92 & 0.92 \\
\textbf{BCCT} & 0.75 & 0.76 & 0.81 & 0.90 & 0.91 \\
\bottomrule
\end{tabular}
\end{table}

As expected, the results show a monotonic gain in all metrics as the population size increases.
Beyond an initial population size of 25 we did not observe significant gains, with both MSCT and LCCT maintaining the same value of 0.87 and 0.92 respectively.
These results show that increasing the population size past 25 test suites does not contribute to the overall performance.

\subsubsection{Ablation Studies}

\begin{table}[ht]
\centering
\caption{Ablation study evaluating the contribution of evolutionary optimization, prompt diversity, temperature diversity, and plateau recovery in EvoGPT.}
\renewcommand{\arraystretch}{1.2}
\resizebox{0.9\columnwidth}{!}{
\begin{tabular}{lccc}
\toprule
\textbf{Configuration} & \textbf{LCCT} & \textbf{BCCT} & \textbf{MSCT} \\
\midrule
LLM-only (no EA) & 83.4 & 80.8 & 80.1 \\
+ EA & 84.9 & 82.1 & 80.8 \\
+ Temperature diversity & 86.2 & 84.6 & 81.3 \\
+ Prompt diversity & 86.8 & 85.5 & 82.0 \\
+ Plateau recovery & 90.0 & 89.0 & 85.0 \\ 
\textbf{EvoGPT (Full)} & \textbf{92.0} & \textbf{90.0} & \textbf{87.0} \\
\bottomrule
\end{tabular}
}
\label{tab:ablation}
\end{table}

To quantify the impact of the core components in EvoGPT, we evaluated multiple ablated
variants in which individual mechanisms were disabled. All ablation experiments were
conducted on the full set of Defects4J projects used in the main evaluation. 
Table~\ref{tab:ablation} presents the performance over the evaluated metrics for each configuration. 
Specifically we compared the following configurations:
\begin{itemize}
    \item \textbf{LLM-only.} Only using the initial population generation component of EvoGPT without temperature or prompt diversity and without the entire population optimization component.
    \item \textbf{+EA.} As above, but also using the population optimization component. Fixed temperatures and prompts, and no plateau monitoring. 
    \item \textbf{+Temperature diversity.} As above, but also using  different temperature settings (but the same prompt) when generating the initial population to add diversity.
    \item \textbf{+Prompt diversity.}
    As above, but also using different prompts to add more diversity.
    \item \textbf{+Plateau escaping.}
    As above, but also using plateau monitoring and escaping. Using a fixed prompt and temperature setting for the plateau escaping strategy.
    \item \textbf{EvoGPT (Full).} The full EvoGPT system, which includes all the above and diverse temperature and prompt settings for the plateau escaping strategy.
\end{itemize}


Overall, the results in Table~\ref{tab:ablation} confirm that each component of EvoGPT contributes incrementally, and that their combined
effect is necessary to fully realize EvoGPT’s performance gains. The results show that each EvoGPT component separately adds less than 4\% to the performance in each of our performance metrics. But the full EvoGPT configuration, which combines an EA to optimize the test population, prompt diversity, temperature diversity, and prompt-diverse plateau recovery, achieves the highest scores across all metrics with an increase of 8.6\%, 9.2\%, and 6.9\% in LCCT, BCCT, and MSCT, respectively. 

In contrast, adding a standard EA population optimization process to LLM-generated tests (the ``+EA'' configuration) provides a limited increase in performance, from  83.4\%, 80.8\%, and 80.1\%,
to 84.9\%, 82.1\%, and 80.8\%, for the LCCT, BCCT, and MSCT metrics respectively.
This result indicates that naively integrating LLM-generated tests and EA does not significantly contribute to the overall performance.


Adding temperature diversity by itself already adds a non-negligible increase in performance compared to using fixed configuration. 
Adding prompt diversity adds another improvement across all metrics, demonstrating that
specialized prompt strategies contribute complementary test behaviors that strengthen the
search process.
Incorporating plateau recovery substantially improves performance by enabling the search to escape local optima through targeted test injection. 
Finally, the full EvoGPT configuration provides an additional significant boost to performance (2\% increase in LCCT and MSCT). This highlights that using diverse prompts also for generating tests to escape local optima is important.


\subsection{Diversity of the Initial LLM-Generated Test Population}
A core component of EvoGPT is the use of five distinct LLM configurations, 
each combining a specialized prompt strategy with an appropriate temperature value, to generate an initial population of 25 test suites. 
This design is intended to generate a more diverse test suite when generating tests using the same prompt and temperature values. 
To demonstrate that indeed more diverse tests are being created in this way, we designed a \emph{diversity analysis} that measures the similarity between the generated tests.
In more detail, we quantified the \textit{semantic differences} in control flow, data usage, and assertion structure, while remaining insensitive to superficial variations such as identifier names or formatting, as follows. 
First, we parse each test suite using \texttt{javalang} and extract semantic features including method declarations, assertion types (e.g., \texttt{assertEquals}, \texttt{assertThrows}), literals (e.g., empty strings, null values, min/max ints, empty arrays, single-element lists), method invocations, and control flow constructs (try-catch blocks, conditionals, loops). 
Then, we used the Jaccard coefficient, a well-established set similarity metric. 
The Jaccard similarity $J(A, B) = |A \cap B| / |A \cup B|$ quantifies the proportion of shared semantic features between test suites, where $|A \cap B|$ is the number of shared features and $|A \cup B|$ is the 
total number of unique features across both suites.

Intra-configuration similarity (average pairwise similarity between test suites from the same configuration) yielded an average of 0.526, whereas inter-configuration similarity (average pairwise similarity between test suites from different configurations) yielded an average score of 0.476. The lower inter-configuration similarity demonstrates that different prompt-temperature configurations produced test suites that are semantically distinct rather than syntactic variants of the same behavior.
This diversity is essential for population optimization process in EvoGPT, since a diverse initial population mitigates premature convergence and enhances the likelihood of covering distinct execution paths and mutants during evolution.

\subsection{Token Counts, Costs, and Runtimes}

\begin{table}[ht]
\centering
\caption{Monetary cost and average runtime per class for each test generation system.}
\begin{tabular}{lccc}
\toprule
 & \textbf{EvoGPT} & \textbf{TestART} & \textbf{EvoSuite} \\
\midrule
Monetary Cost (USD) & 0.32 & 0.01 & 0.0 \\
Avg. Runtime (min.)   & 8  & 2  & 1 \\
\bottomrule
\end{tabular}
\label{tab:cost-runtime}
\end{table}


Table~\ref{tab:cost-runtime} reports the total monetary cost and average runtime \textit{per class} observed in our experiments with EvoGPT, TestART, and EvoSuite.
Running EvoGPT incurred a cost of \$0.32 USD per class across all projects, with an average runtime of 8 minutes per class, 
while TestART incurred a total cost of \$0.01 USD with an average runtime of 2 minutes per class. 
EvoSuite, which does not call remote APIs and is executed locally, incurred no monetary cost and had an average runtime of 1 minute per class on average. 
As expected, EvoGPT is more expensive than the baselines in both runtime and token usage. Yet it still remains within a manageable range for many software engineering use cases. 
Our experiments, conducted under a fixed evolutionary budget (population and max generations), demonstrate that paying these costs allows EvoGPT to yield significantly better test suites, effectively trading computational time and monetary costs for test quality. 



\subsection{Limitations}
While EvoGPT generates impressive results across the Defects4J benchmark, it suffers from the following limitations. 

\noindent \textbf{Runtime.} As shown in Table~\ref{tab:cost-runtime}, EvoGPT requires significant computational cost. Although the initial population generation is executed asynchronously, generating 25 diverse unit test suites using OpenAI's API remains time-consuming, particularly for large and complex classes. 
Furthermore, the evolutionary population optimization process also incurs a significant runtime per generation since computing our fitness function requires instrumenting, compiling, and running each test to perform coverage and mutation analysis. 
As a result, the end-to-end test generation process for a single source class can be slower than the baselines, which may limit scalability in large-scale or time-constrained testing scenarios.

\noindent \textbf{Monetary cost.} Our system relies on LLM-based test generation for generating the initial population of 25 unique test suites and for generating tests to escape plateaus. 
This comes with a tangible monetary cost due to API calls to commercial LLM providers (OpenAI in our case), as seen in Table~\ref{tab:cost-runtime}. 
As such, the per-class cost of test generation may be prohibitive at scale, especially in industrial settings with hundreds or thousands of classes.
However, in such settings one may select an open-source free LLM model to reduce costs.

\noindent \textbf{Evolutionary budget vs. wall-clock time.} To isolate the algorithmic contributions in EvoGPT, we controlled our experiments using a fixed evolutionary budget (generations and population size) rather than wall-clock time. 
With the exception of the plateau escaping step, EvoGPT's runtime per generation is similar to that of EvoSuite. 
However, the initial population generation process of EvoGPT is significantly slower, which may limit its applicability in time-constrained environments, e.g., CI/CD pipelines with strict time restrictions. 
However, for nightly builds or safety-critical validation where coverage is paramount, the overhead of EvoGPT is viable and worth the gains in fault detection. 


\subsection{Threats to validity}
\paragraph{Internal Validity.} Our reported coverage and mutation scores depend on the accuracy of third-party tools: JaCoCo for line/branch coverage, and PITest for mutation analysis. These tools are widely used, but they are not perfect. For example, PITest may not detect equivalent mutants (mutants that behave identically to the original code), which can falsely lower the mutation score. Similarly, JaCoCo may miss certain paths or lines due to bytecode instrumentation quirks. These limitations could slightly distort the results and should be considered when interpreting our metrics.

\paragraph{Construct Validity.} We evaluated generated test suites fitness using structural and mutation-based metrics (LCCT, BCCT, MSCT).
However, we did not measure other qualitative aspects such as readability, assertion relevance, or developer trust.
In addition, EvoGPT leverages temperature-based LLM sampling, which are stochastic. 
Thus, the generated test suites may vary across runs. While our reported results reflect representative averages, future work could analyze variance across seeds to better characterize the stability of the generated test suites.  

\paragraph{External Validity.} Our experiments are limited to Defects4J projects only. Although these span a range of domains, they may not fully represent the complexity of industrial code bases. Furthermore, the focal methods used in our experiments are public methods, which may not capture the full interaction behavior in production scenarios. 

\paragraph{Reproducibility.} 
Our code, data, and scripts are available at \url{https://tinyurl.com/EvoGPT}. 
Exact reproducibility depends on access to specific LLMs and API configurations. Moreover, LLM outputs are stochastic and the LLM we used (gpt-4o-mini) is periodically updated by OpenAI.

\section{Conclusion and Future Work}
\label{sec:conclusion}
In this paper, we introduced \textbf{EvoGPT}, a hybrid automated test generation system
that integrates LLM-based and EA-based test generation techniques. 
EvoGPT uses
multiple prompt and temperature configured LLM agents to generate a semantically diverse
initial population of tests, which is subsequently refined through a population optimization
process guided by coverage and mutation-based objectives.
In addition, EvoGPT employs a prompt-diverse plateau escape mechanism that selectively
re-engages LLM agents when evolutionary progress stagnates, injecting targeted test
methods to escape local optima.
Our empirical evaluation on
Defects4J projects demonstrates that EvoGPT consistently outperforms established
LLM-based and EA-based baselines in terms of line coverage, branch coverage, and
mutation score, highlighting the effectiveness of combining LLM diversity with
search-based refinement. 
Our ablation study also highlights that naive integration of LLM generated test cases and EA optimization yield limited advantage, and that temperature and prompt diversity have significant effect. 

Several directions remain for future work. First, we plan to extend the evaluation to
larger benchmarks and open source projects, including additional programming languages, as well as to assess the bug-detection capability of EvoGPT-generated tests against real historical defects.
Second, EvoGPT could benefit from incorporating grep search in order to fetch imported objects signatures and definition, in order to provide richer contextual information to the LLM component and to further guide the EA. Finally, exploring cost-aware strategies for
adaptive test generation may improve the practicality of EvoGPT in large-scale or
resource-constrained settings.

\bibliographystyle{IEEEtran}
\bibliography{software}

\end{document}